\documentclass[aps,prl,floatfix,twocolumn,groupedaddress,showpacs,longbibliography]{revtex4-1}

\usepackage[utf8]{inputenc}
\usepackage[english]{babel}
\usepackage{float}
\usepackage{amsmath,amssymb,graphicx,enumerate,bm}

\usepackage{color}
\usepackage{hyperref}

\begin{document}

\title{Zero gravity thermal convection in granular gases}
\author{A. Rodr\'iguez-Rivas}
\author{M. A. L\'opez-Casta\~no}
\author{F. Vega Reyes}
\affiliation{
Departamento de F\'{\i}sica and Instituto de Computaci\'on Cient\'{\i}fica Avanzada (ICCAEx), Universidad de Extremadura, 06006 Badajoz, Spain \\
}

\date{\today}

\begin{abstract}
  Previous experimental and theoretical evidences have shown that convective
  flow   may appear in  granular fluids, if subjected to a thermal gradient and  gravity (Rayleigh-B\'enard type convection).  In contrast to
  this, we present here evidence of gravity-free thermal convection in a granular gas, with no presence of external thermal gradients either.
 Convection is here maintained steady by internal gradients due to dissipation and thermal sources at
  the same temperature. The granular gas is composed
  by identical disks and is enclosed in a rectangular region. Our results are
  obtained by means of an event driven  algorithm for inelastic hard disks.
\end{abstract}

\maketitle

Granular dynamics has been an interesting test ground in the last decades for
non-equilibrium statistical mechanics and complex fluid mechanics
\cite{JNB96,AT06}. We know from previous works that much of the phenomenology
observed in molecular gases and condensed matter \cite{dG92} arises in granular
matter as well, but in general with added complexity. Phenomena like jamming
\cite{jamming,LN98,LBK08,BISE13}, crystallization \cite{AW62,PSP83,SL96}, glass
transitions \cite{glass,PH13}, capillarity \cite{FPP17}, fluid flow and convection \cite{CH93,VU09}, memory
effects \cite{KAHR79,LVPS19} etc., appear also in granular matter systems. But,
furthermore, there is also a rich phenomenology which is intrinsic to granular media,
such as clustering instabilities in low density systems \cite{GZ93,MIMA08} or
inelastic collapse in denser systems \cite{OU98}.

Granular convection and pattern formation in systems under gravity has been
known for quite some time now \cite{F31,JNB96,VIP}. It has been observed for
instance in experiments with vertically oscillated particles
\cite{UMS96,BSSMS98}. In the case of horizontally unbounded low density systems,
previous works have provided complete descriptions of the different types of
patterns that can be observed by means of computer simulations
\cite{BSBSMS98,BSSMS98,RRC00} and theoretical studies \cite{KM03,NAAJS99}. These
works have shown the existence of a formal analogy with the classical
Rayleigh-B\'enard convection in molecular fluids \cite{CH93}. However,
experimental work shows, to a certain degree, a mismatch with the theory; for
instance, in the threshold values of the buoyancy driven convection
\cite{EWMBL07}. Part of the origin of this disagreement stems from the existence
of another type of convection mechanism due to dissipation at the sidewalls. And
of course, sidewalls are inherently present in granular dynamics experiments. In
fact, the presence of sidewalls is known to have impact on hydrodynamic
instabilities in general, even if they are physically inert
\cite{CH93,RRC00,PVWT14}. More specifically, thermal convection induced by
sidewall energy leaks is well known in molecular fluids \cite{HW77}. For
granular materials, a similar mechanism is triggered by wall-particle inelastic
collisions, rather than a thermal leak \cite{WHP01,TV02,PGVP16,WLMP18}.

In any
case, and to our knowledge, thermal convection in granular dynamics has always
been detected in the presence of a gravitational field (see references above and
their bibliographies) \cite{yules_exception}. We now prove the existence of
zero-gravity granular thermal convection \cite{vip_exception}. Furthermore, as
we will show, the intrinsic thermal gradient (induced by particle-particle
inelastic collisions \cite{PGVP16}) is not strictly necessary to produce
convection, requiring only sidewalls energy dissipation (and the presence of
thermal walls, that set the steady state temperature value). This result obviously
has an impact in granular matter applications under microgravity or no gravity conditions (involving
mining, storage and transportation of granular matter), and in experiments in
the near future by space agencies \cite{S19}. For instance, an experimental evidence of the new
convection is foreseeable in the context of research projects in low-gravity or no
gravity environments \cite{VIP}.

\begin{figure}
    \centering
    \includegraphics[width=.6 \columnwidth]{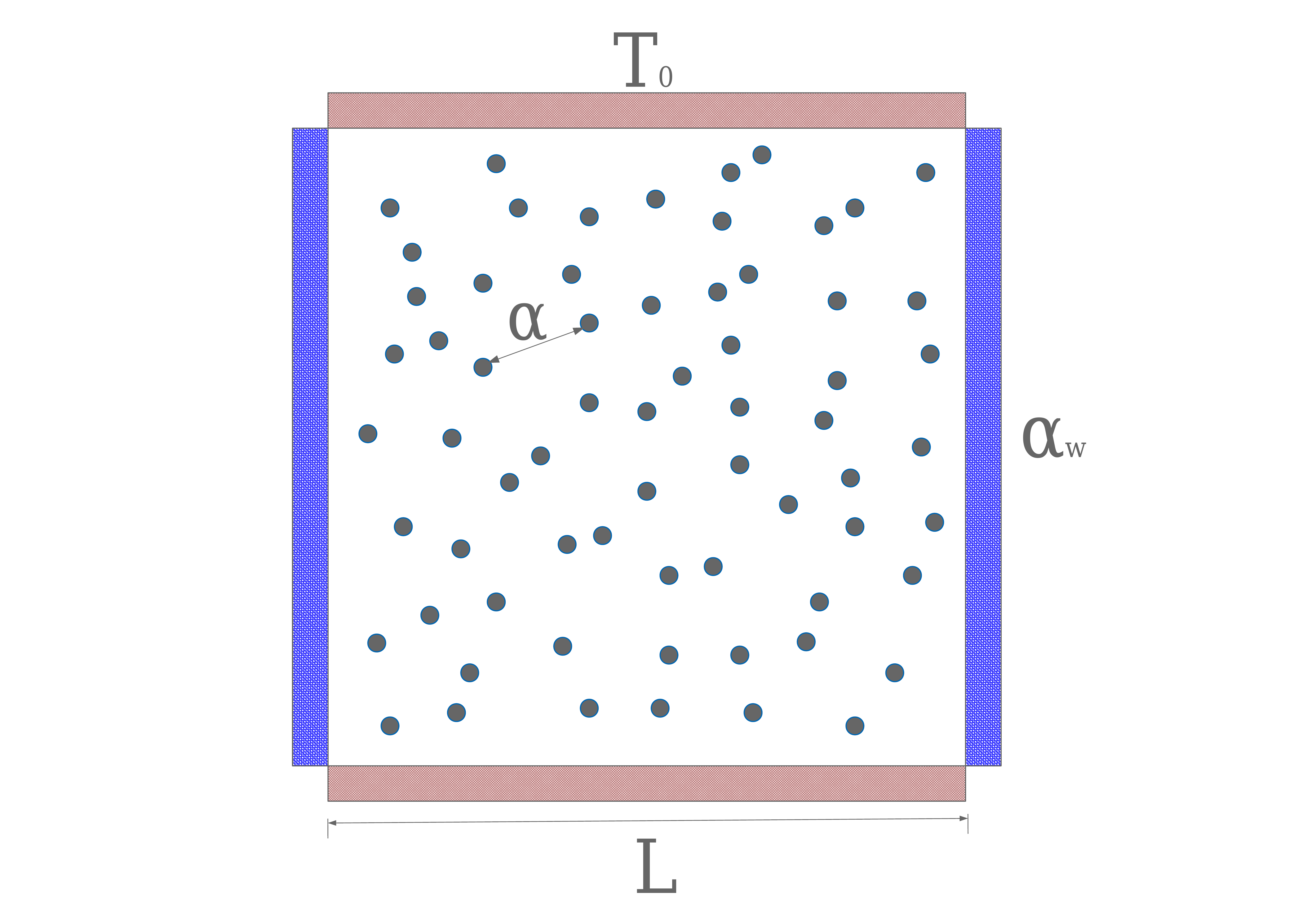}
    \caption{Sketch of the system. The only thermal sources are at the upper and bottom walls
      (both set at the same temperature $T_0$, thus injecting energy to
      particles nearby through a standard computational procedure
      \cite{supplemental}), and without inducing temperature difference between the
      fluid near both walls. 
    The other two parallel walls (left and right) are inert and dissipative; 
    i.e. particles undergo inelastic collisions at contact with these walls. } 
    \label{sketch}
\end{figure}

\begin{figure}[!t]
\centering
\includegraphics[width=.85 \columnwidth]{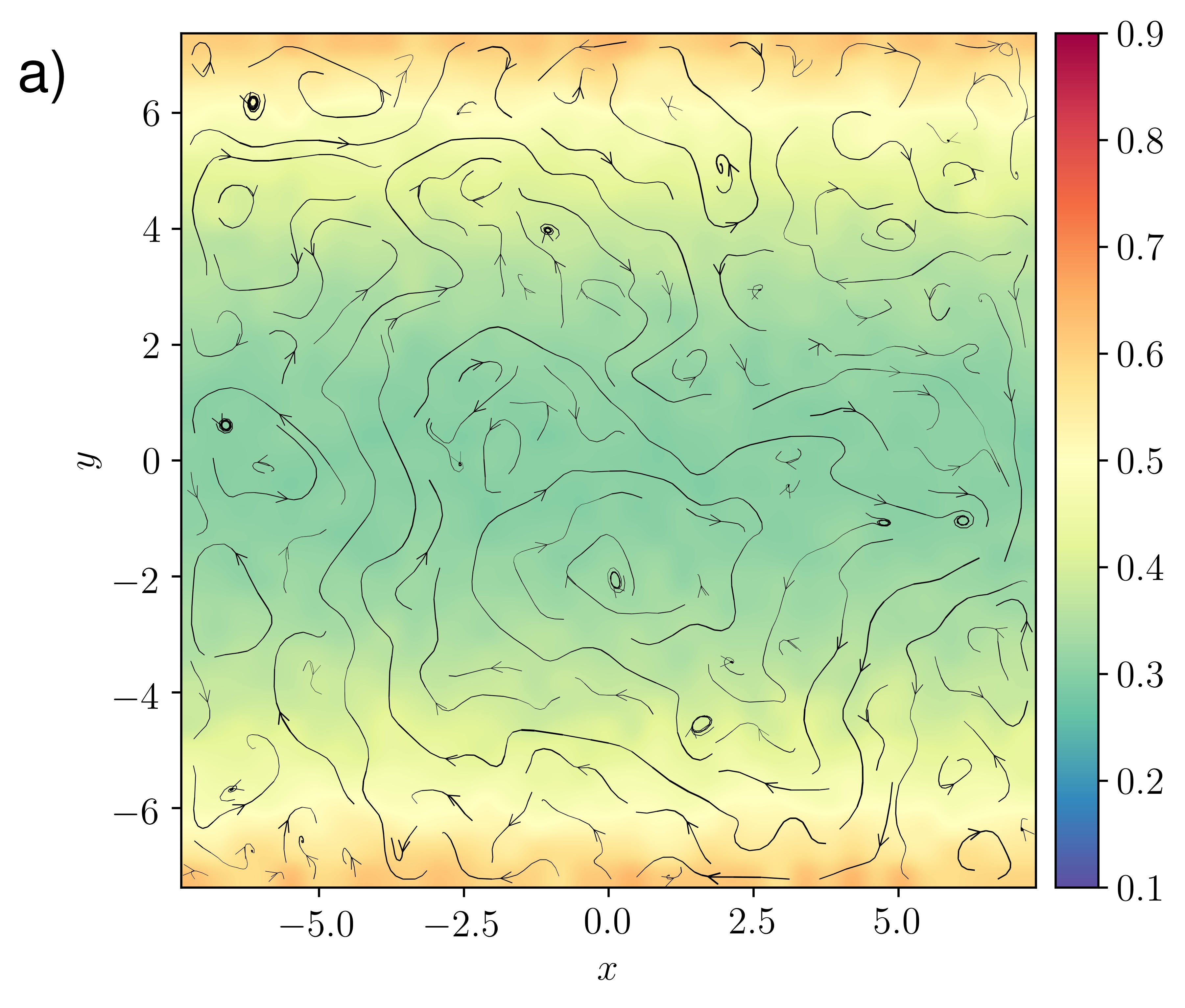}
\includegraphics[width=.85 \columnwidth]{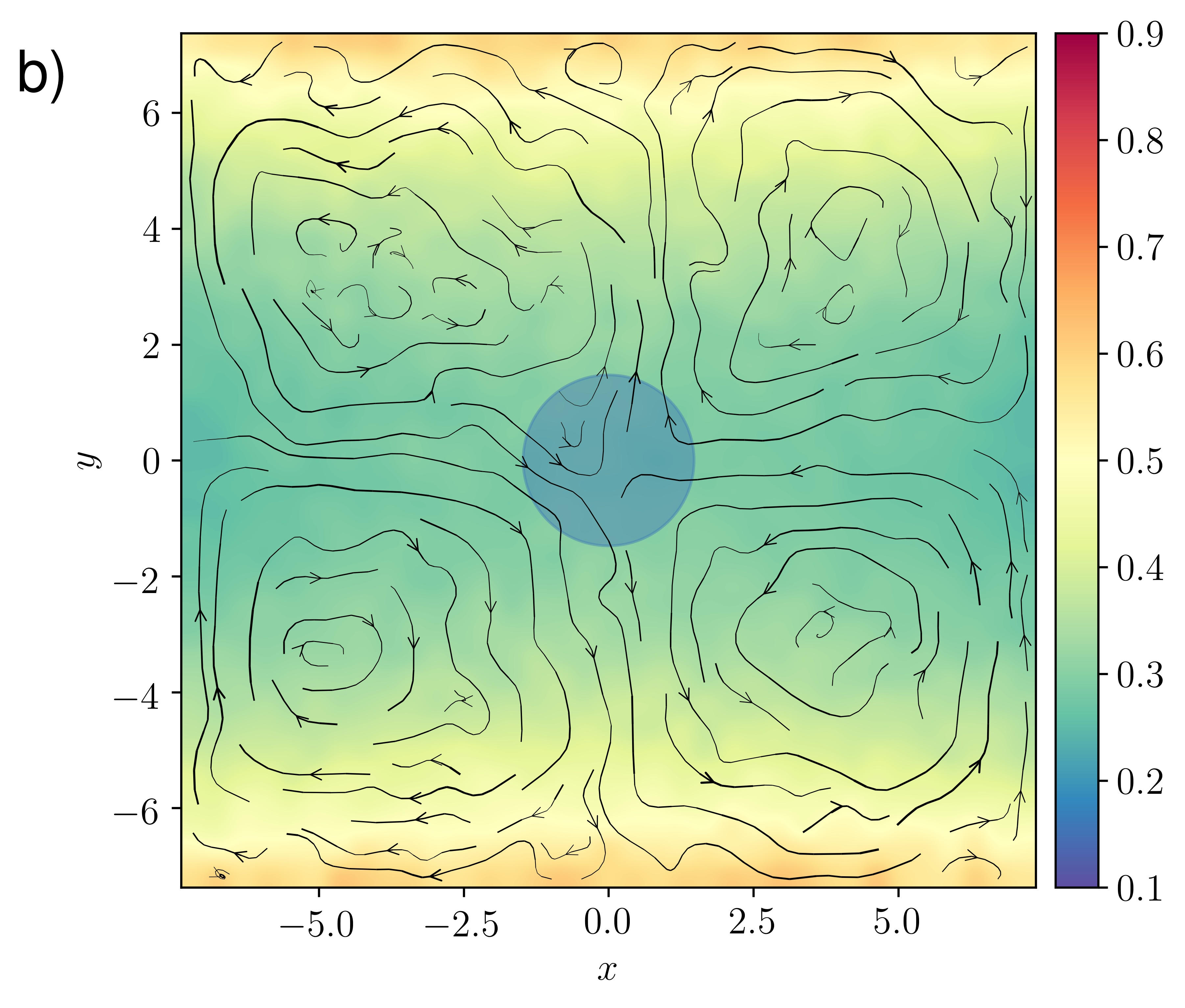}
\includegraphics[width=.85 \columnwidth]{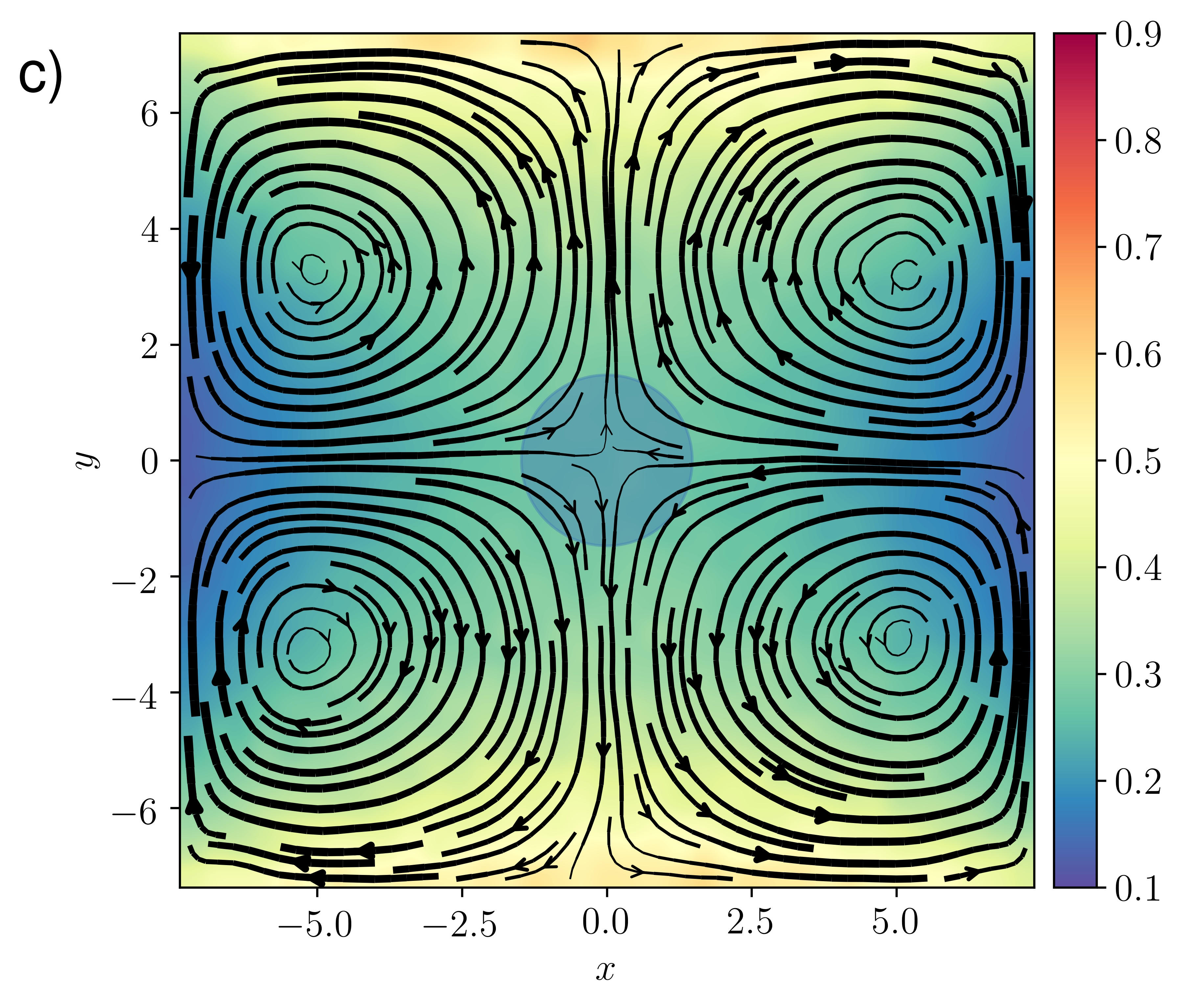}
\caption{Flow velocity ($\mathbf{u}$) and granular temperature ($T$)
  hydrodynamic fields, with $\alpha = 0.9$, $L=15\lambda$, and: (a) $\alpha_w=1$
  (elastic sidewalls, no convection), with
  $u_{\mathrm{max}}=\left(\sqrt{u_{x}^{2}+u_{y}^{2}}\right)_{\mathrm{max}} =
  0.00272$; (b) $\alpha_w=0.9$ (incipient convection), with
  $u_{\mathrm{max}}=0.00363$; (c) $\alpha_w=0.6$ (fully developed convection),
  with $u_{\mathrm{max}}=0.04008$. Line thickness is relative to flow
 intensity. Flow centers appear highlighted in clear blue.} \label{lp1_TV}
\end{figure}

Let us describe our system in more detail. We deal with a 2D granular gas
enclosed in a square-shaped system, as sketched in Figure
\ref{sketch}. Particles are identical smooth hard disks with mass $m$ and
diameter $\sigma$, used as mass and length units in this work. Particle
density $n$ remains sufficiently low at all times in the system, so that collisions
are always binary and instantaneous. Since we use the smooth hard particle
collisional model, rotational degrees of freedom are neglected \cite{KM03}. Coefficients of normal
restitution $\alpha$ and $\alpha_{w}$ characterize the degree of inelasticity
upon particle-particle and sidewall-particle collisions respectively \cite{PGVP16}. A
pair of thermal walls injects kinetic energy to nearby particles.  We label the
other two walls as ’lateral’ walls or sidewalls. The system is free of any gravitational field, thus avoiding any possibility of buoyancy driven convection.

As seen in a previous work \cite{PGVP16}, if a system like the one in
Figure~\ref{sketch} is under the action of gravity, there is no mathematical solution
to the corresponding hydrodynamic equations for a hydrostatic state,
$\mathbf{u}=\mathbf{0}$. Therefore, the action of gravity combined with dissipative
sidewalls leads automatically to convection (this happens in molecular fluids as well
\cite{HW77}). But this is not so in the absence of gravity.

In effect, by taking into account the symmetry properties for the system in the
steady state ($\partial/\partial t=0$), we can obtain the corresponding equations for
a hydrostatic state \cite{supplemental} by imposing the condition $\mathbf{u}=\mathbf{0}$ on the granular
gas balance equations. For disks we obtain the following differential equation for the
temperature \cite{VU09}

\begin{align}
& \frac{\sqrt{T}}{\pi\sigma^2}\left[\frac{\partial}{\partial x}\left(\sqrt{T}\frac{\partial T}{\partial
  x}\right) + \frac{\partial}{\partial y}\left(\sqrt{T}\frac{\partial
  T}{\partial y}\right)\right]=\frac{\zeta^*(\alpha)}{\kappa^*(\alpha)}p^2\: , 
\label{2DT}
\end{align}
which now admit a non-trivial solution. Hence, the base state (the simplest
hydrodynamic state) for $g=0$ is a hydrostatic one. 
In \eqref{2DT}, $p$ is the hydrostatic pressure,
$T$ is the granular temperature, 
$\kappa^*(\alpha)$ is the transport coefficient associated to the
heat flux and $\zeta^*(\alpha)$ is the cooling rate due to particle-particle
inelastic collisions \cite{BC01}. This means that an eventual gravity-free convection would appear
only under certain conditions.


\begin{figure*}[!t]
\centering
\includegraphics[width=.85 \columnwidth]{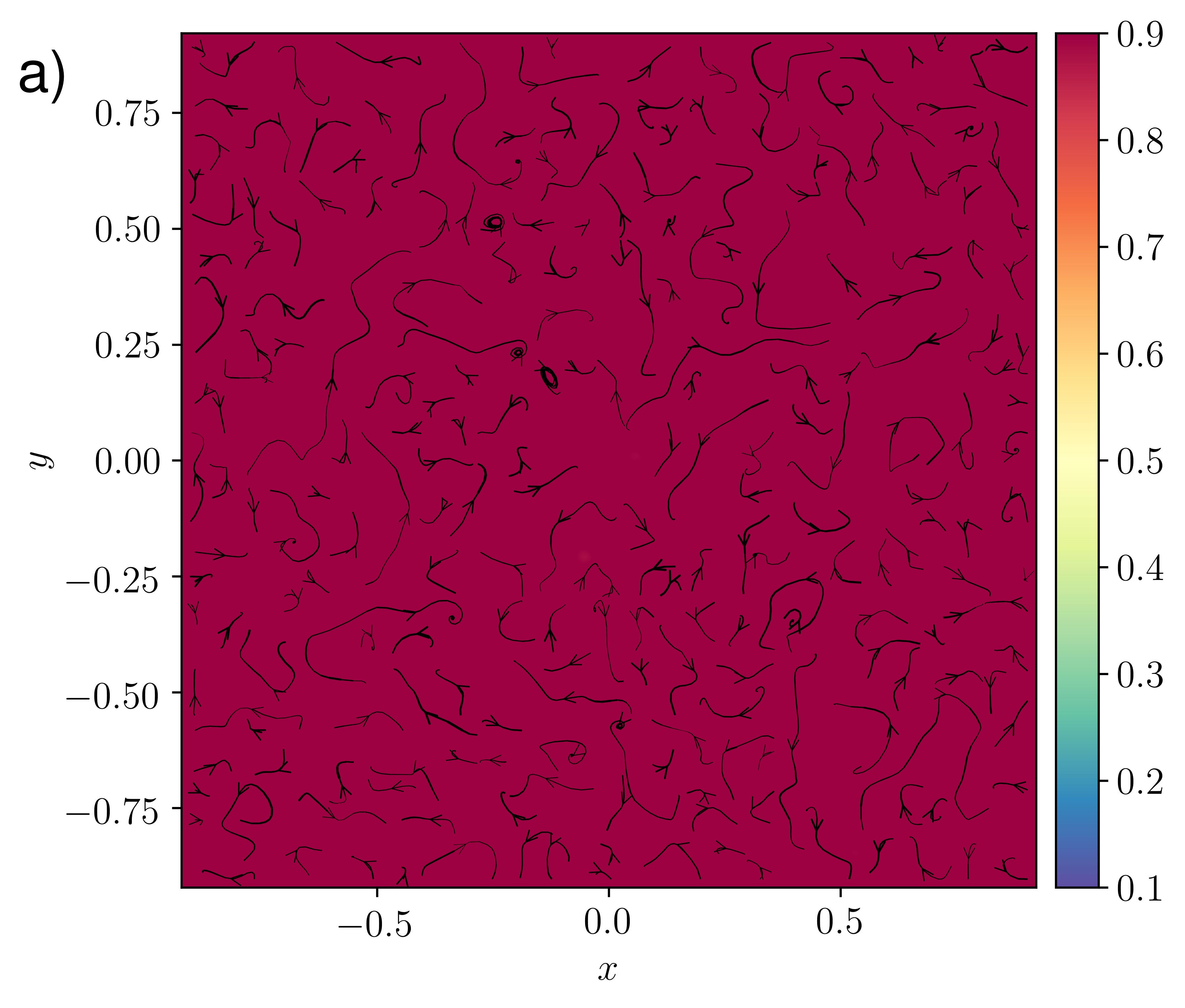}\quad
\includegraphics[width=.85 \columnwidth]{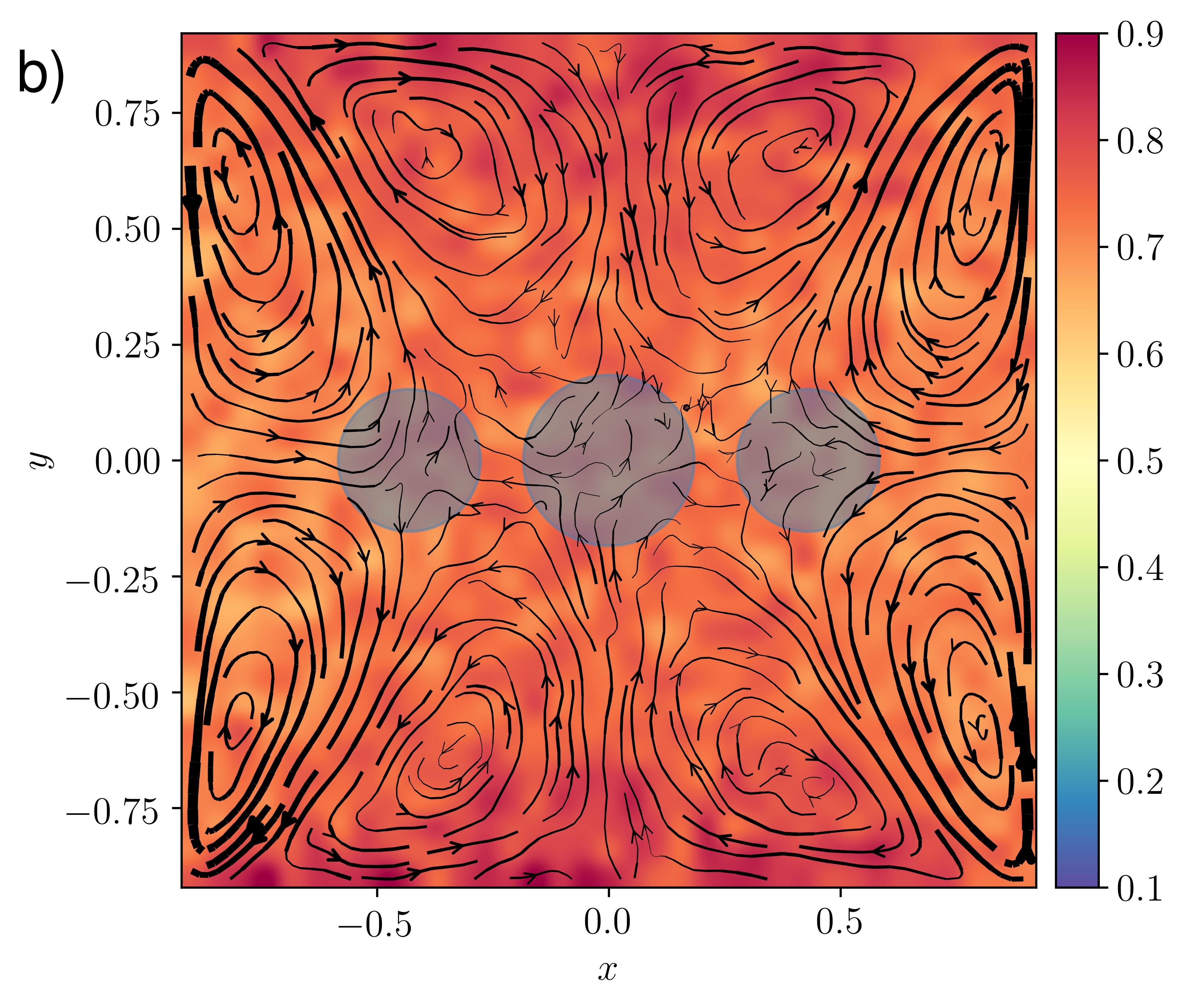}
\includegraphics[width=.85 \columnwidth]{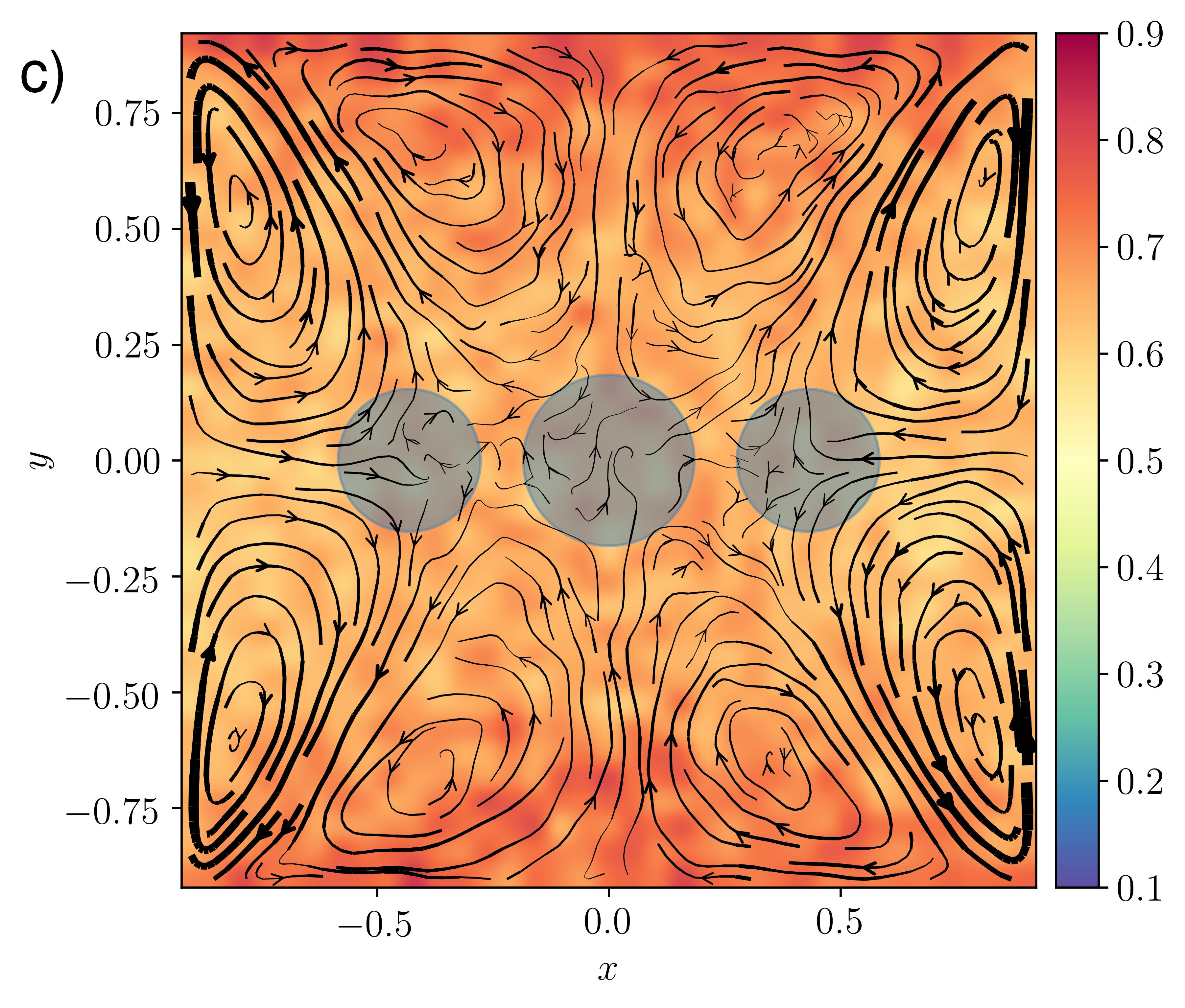}\quad
\includegraphics[width=.85 \columnwidth]{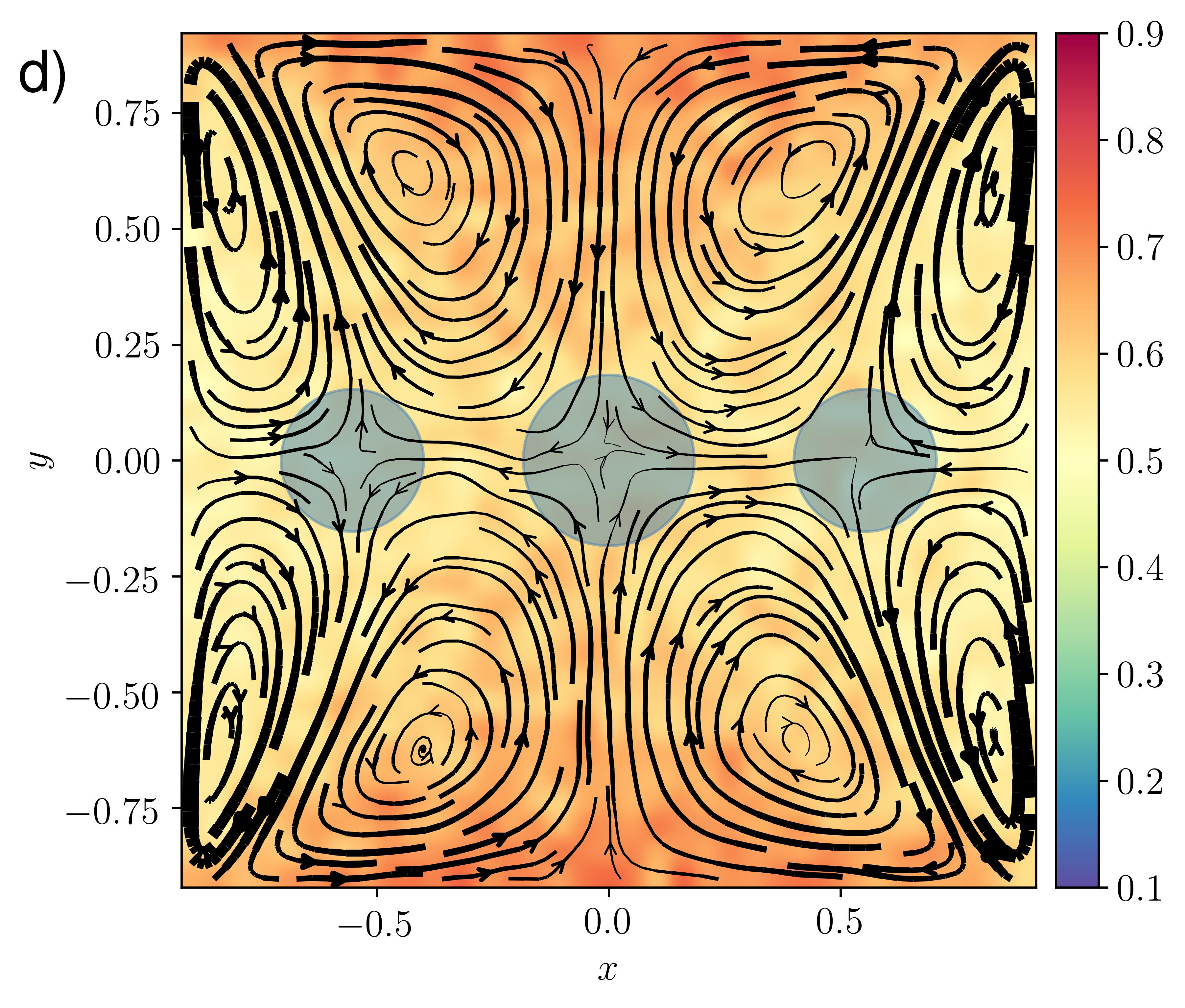}
\caption{Hydrodynamic fields $\mathbf{u}$ and $T$ for $L=(15/8)\lambda$, with: (a)
  $\alpha=1.0$ $\alpha_{w}=1.0$; $u_{\mathrm{max}}=0.00450$, (b) $\alpha=1.0$,
  $\alpha_{w}=0.7$; $u_{\mathrm{max}}=0.02453$, (c) $\alpha=0.9$, $\alpha_{w}= 0.7$;
  $u_{\mathrm{max}}=0.02254$, (d) $\alpha=0.9$, $\alpha_{w}=0.5$;
  $u_{\mathrm{max}}=0.03686$. Flow centers appear highlighted in clear blue.}  
\label{lp8_TV}
\end{figure*}

In order to analyze the problem, we perform event driven simulations of inelastic
smooth hard disks \cite{supplemental,DYNAMO}. 
Throughout this work, results are represented in dimensionless variables (denoted
with the same symbols as their dimensional counterparts), by using $\sigma$, $\tau = \sqrt{m
  \sigma^{2} /T_{0}}$, $m$ and $T_{0}$ as units of length, time, mass and
temperature, respectively; where $T_{0}$ is the thermal walls temperature. In order
to characterize particle density, we use the two-dimensional packing fraction $\bar{\phi}=N\pi\sigma^{2}/4L^{2}=10^{-3}$, being $N$ the total number of disks in the system. Therefore the mean free path is $\lambda=\left(2\sqrt{\pi}\bar{n}\sigma \right)^{-1}=221.5567$,  where $\bar{n} = N/L^{2}$ is the system average particle density.
The specifics of the simulation are discussed in the Supplemental Material file \cite{supplemental}.




Figure \ref{lp1_TV} displays flow velocity ($\mathbf{u}$) and granular
temperature ($T$) fields for a system with $L= 15\lambda$ and $\alpha =0.9$, for
three different $\alpha_{w}$ values ($\alpha_{w} = 1, 0.9 ~\mathrm{and}~
0.6$). As it can be seen, convection is initially absent for $\alpha_{w} = 1$ (only remnant noise is observed)
but develops for $\alpha_{w} \le \alpha_{w}^{\mathrm{th}}(\alpha) < 1$, where
$\alpha_{w}^{\mathrm{th}}(\alpha)$ is the convection critical value, that depends on
$\alpha$ and the other system parameters. Moreover, convection is strongly dependent on the parameter $\alpha_w$, this
being possibly due to the strong correlation between sidewalls dissipation and
the temperature gradient \cite{supplemental}. 
Notice that, in the presence of gravity, sidewall dissipation generates just one
cell attached to each dissipative wall \cite{PGVP16}, whereas now we find 2
convective cells per dissipative wall. This result makes sense since, for our
geometry and with $g=0$, streamlines should have two perpendicular axes of specular
symmetry, both passing through the center of the system. 


Streamlines coming from dissipative sidewalls flow towards the center, then bending
off the center onto the thermal walls.  
In fact, the orientation of the flow in the convection cells can be explained by
looking at the behaviour in certain singular points, where streamlines in
convection rolls meet. We mark these \textit{flow centers} with circles in
Figures~\ref{lp1_TV} and~\ref{lp8_TV}. In this sense, we can see that in Figure
\ref{lp1_TV} there is only one flow center, which coincides with the center of
the system whereas in Figure~\ref{lp8_TV}, we find 3 flow centers: one in the system
center and other two closer (and at the same distance) to the sidewalls. Notice also
that these two flow centers approach the sidewalls as they become more inelastic
(Figures 3c and 3d), which can be an indication that the convection mechanism is related to the
inability of colder particles to reach the system center, as they flow from the sidewalls.





In Figure \ref{lp8_TV} we show the results for a smaller system ($L =
(15/8)\lambda$). As we see, the convection pattern is now very different. We now
observe 4 cells next to dissipative walls plus 4 additional cells emerging near the
system center, totalling 8 convection cells and three flow centers.  
Also notice that decreasing the value of $\alpha_w$ expands the area of the four
central cells while decreasing the size of the cells next to the lateral walls. A
reason for the appearance of the new central cells may be that the center is now
hotter than in the bigger system in Figure~\ref{lp1_TV} \cite{VU09}. This produces a new
convection center from which streamlines flow out. 

It is very important to notice (Figure 3b) that convection can appear even for
elastic particle-particle collisions. As we see, a 2D thermal gradient, created
only by dissipation at the lateral walls ($\alpha_w<1$) and the thermal sources
in the upper and lower walls is sufficient to trigger convection. Figures \ref{lp1_TV}, \ref{lp8_TV} show a general trend of stronger convection for overall increasing inelasticity, this trend being more important for
sidewall dissipation increase (decrease of $\alpha_w$).

\begin{figure}[ht]
\includegraphics[width=.85 \columnwidth]{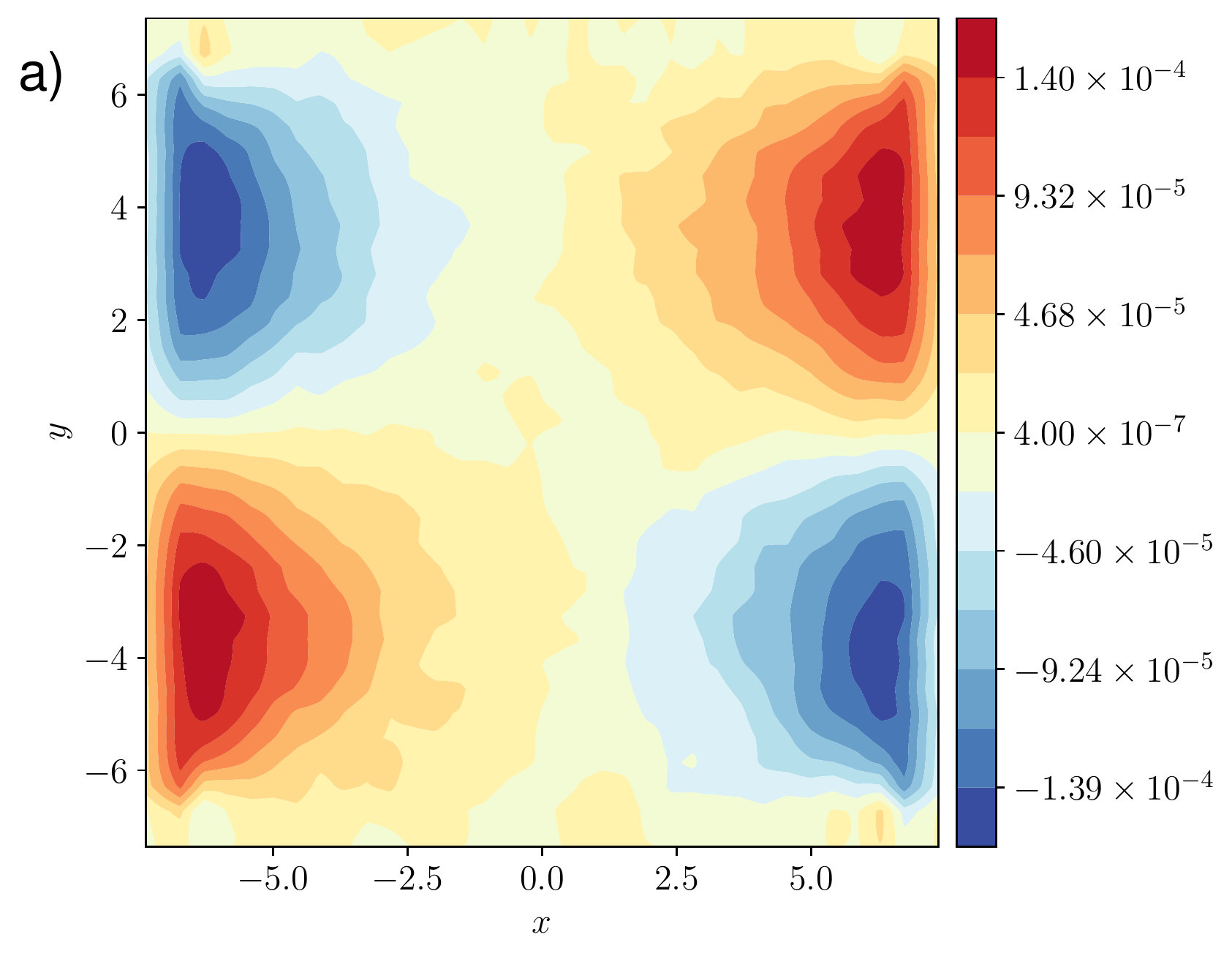}
\includegraphics[width=.85 \columnwidth]{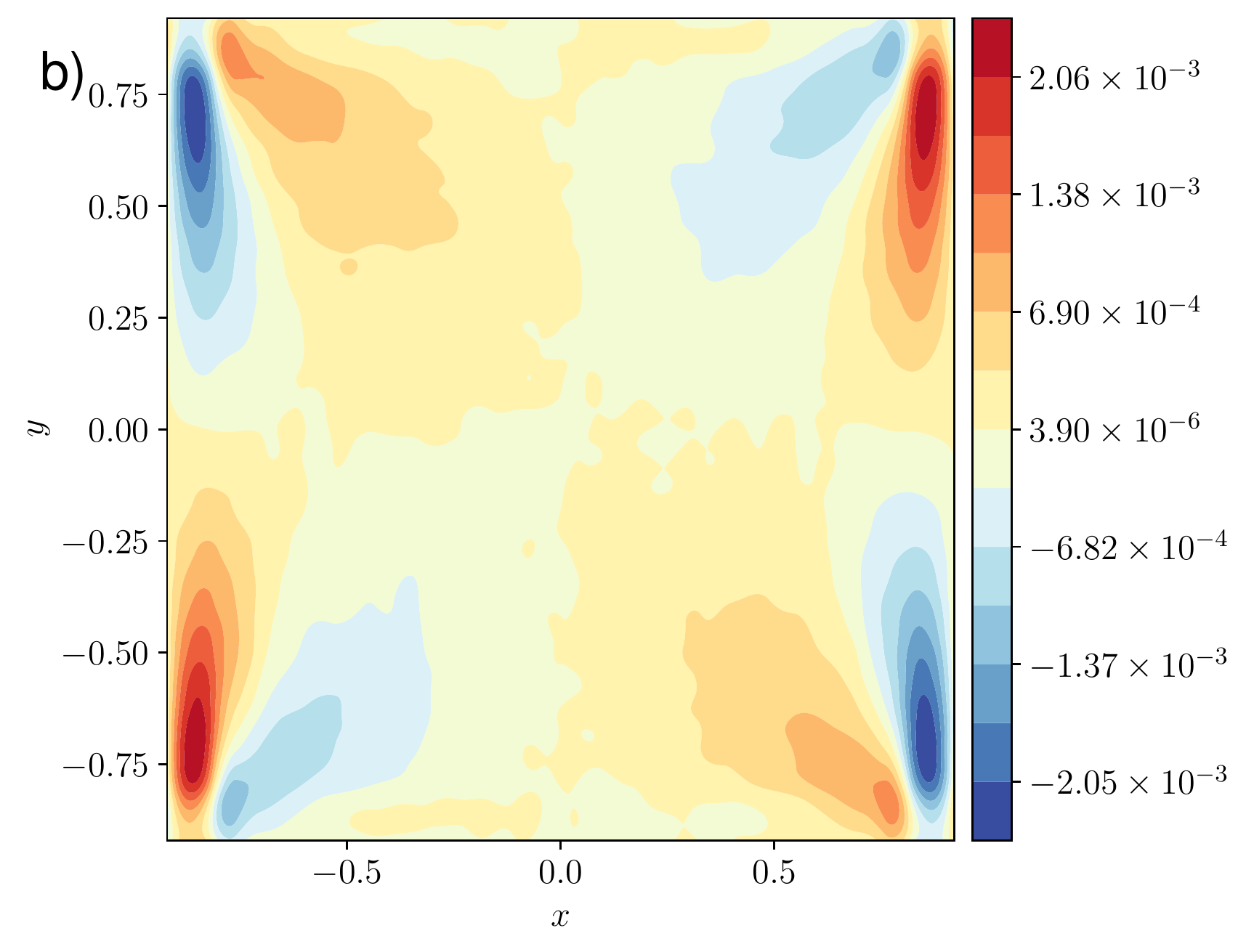}
\caption{Vorticity field $\omega-\omega_0$ for $\alpha=0.9$ and $\alpha_w=0.4$, 
obtained for the box sizes (a) and $L=15\lambda$, (b) $L=(15/8)\lambda$. $\omega_0$ is the base vorticity level
coming from noisy data at $\alpha=\alpha_w=1$.} \label{V_local_cell}
\end{figure}

\begin{figure}[ht]
\includegraphics[width=.85 \columnwidth]{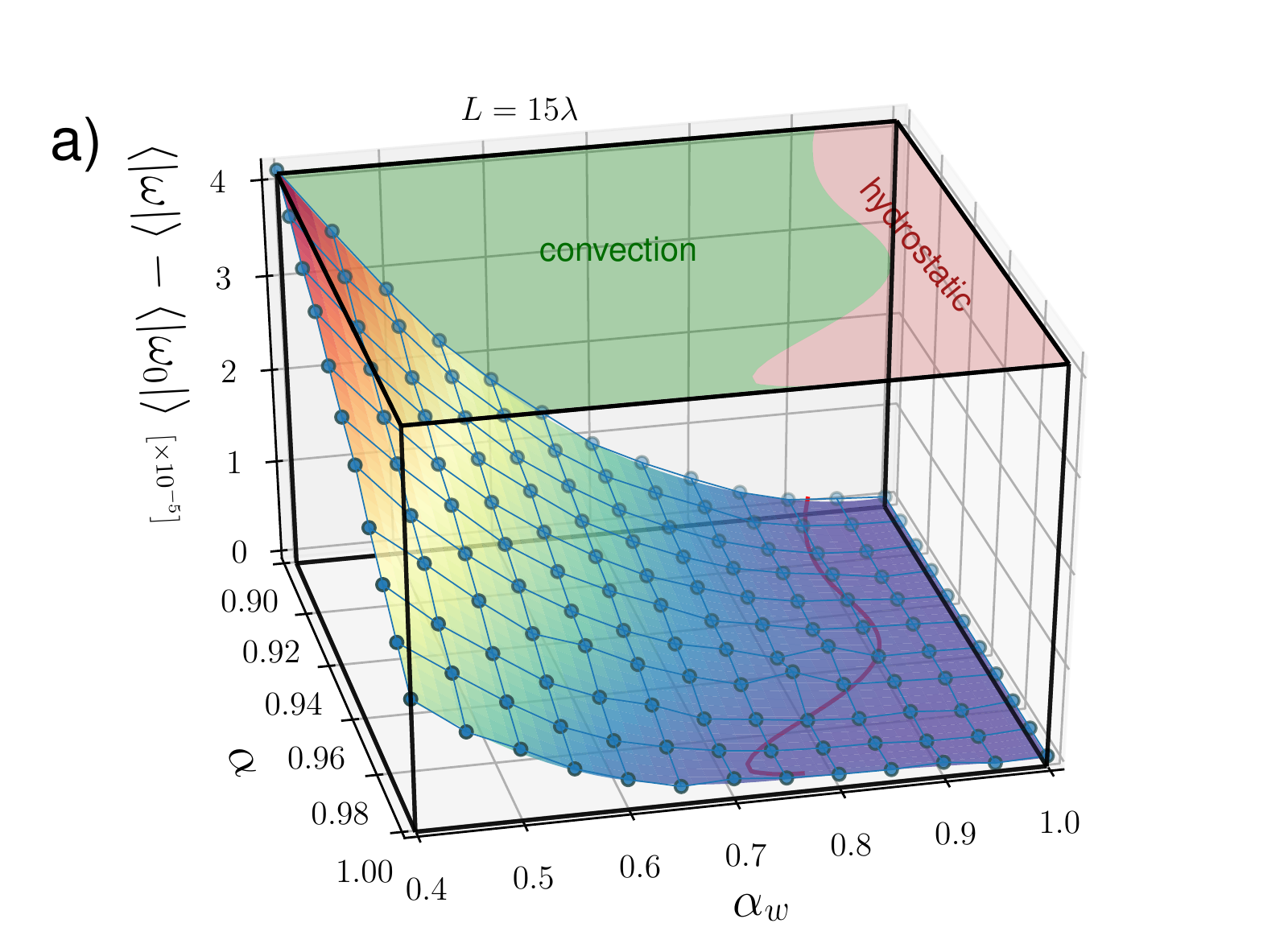}
\includegraphics[width=.85 \columnwidth]{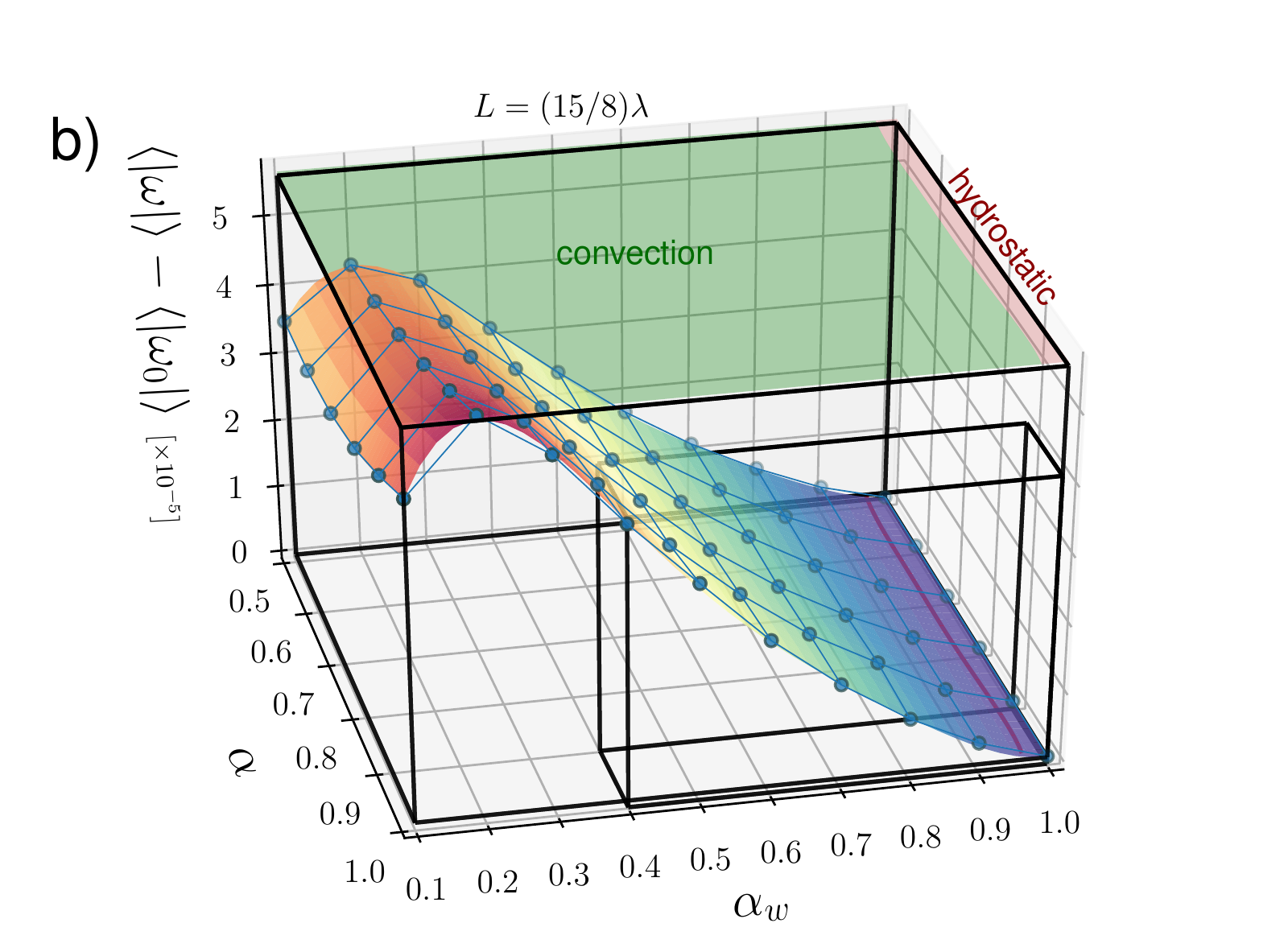}

\caption{Global average (i.e., averaged over all points in the system) of the vorticity absolute value, $\langle|\omega|\rangle-\langle|\omega_0|\rangle$ against $\alpha$ and $\alpha_{w}$. (a) For $L=15\lambda$ and (b) For
$L=(15/8)\lambda$. A diagram showing areas where convection/hydrostatic states is
projected on the top of each figure. The curve separating both regions marks the
$\alpha_w^\mathrm{th}$($\alpha)$ critical values. The smaller framing box in (b) marks the
parameter space represented in (a). $\langle|\omega_0|\rangle$ is the average base vorticity absolute value
coming from noisy data at $\alpha=\alpha_w=1$.}
\label{V_cell}
\end{figure}

A convenient way to analyze the convection intensity is by looking at the
vorticity field, $\omega \equiv \displaystyle{\partial_x u_{y} - \partial_y
  u_{x}}$, as in Figure \ref{V_local_cell}. Two distinctive convection patterns
are found, as already mentioned, with either 4 or 8 cells with alternating
vorticity sign, the 4 central cells disappearing for bigger systems. Notice
that, in Figure \ref{V_local_cell}b, vorticity nuclei are closer to the
corners. As we commented above, an important aspect of gravity-free convection
is that the system allows for hydrostatic states and hence convection is not
expected to appear in all cases. Figure \ref{V_cell} illustrates this point,
where convection threshold lines are shown on the global vorticity surface $\langle|\omega|-|\omega_0|\rangle(\alpha$, $\alpha_w)$. We clearly detect non-convective
(hydrostatic) regions. Moreover, the hydrostatic
state tends to occupy wider regions in the parameter space as the system increases in
size. Vorticity surface reveals clearly (Figure~\ref{V_cell} b) that,
although gravity-free convection is indeed
stronger for more inelastic systems, the opposite trend is observed near a region
close to $\alpha_w=0.2$, where a significant vortiticty drop towards $\alpha_w\to 0$
is observed.

In summary, in this work we have shown the existence of gravity-free granular convection. It is
produced by the existence of sufficiently strong a 2D thermal gradient out of an initially hydrostatic base
state at small gradients. Since 2D thermal gradients are actually characteristic of granular experiments
(usually being produced at system boundary corners), this
type of convection should be present in most zero gravity experiments. From a
fundamental point of view, it is interesting to remark that molecular gases
should also display gravity-free convection as long as they present analogous boundaries to those studied here. In
fact, analogous gravity-free natural convection may be found in liquid droplets \cite{KMSAS17, PSK18}  due to
thermocapillary effects driven by the gradients of surface tension, the difference being here that
now the 2D thermal gradient is generated by the boundary curvature. From a more
general point view, our
results constitute a rare example of natural thermal convection with no gravity,
in a generic fluid system. Moreover, it would be interesting to test if other effects such as capillarity \cite{FPP17} also appear under no gravity conditions.
Additionally, we think the results in this work may have an impact in other contexts such as horizontal
systems (common in living matter for instance). Finally, the mechanism for the
onset of gravity-free natural convection is outlined.


\begin{acknowledgments}
 We acknowledge funding from the Government of Spain through project No. FIS2016-76359-P and from
 the regional Extremadura Government through projects No. GR18079 \& IB16087, both partially
 funded by the ERDF.
\end{acknowledgments}

\bibliography{G0}

\begin{thebibliography}{43}
\bibitem [{vip()}]{vip_exception}%
  \BibitemOpen
  \href@noop {} {}\bibinfo {note} {As noted in the reference by Auma\^itre et
  al. in our bibliography, even under low-gravity, convection has been detected
  just in a computational study with polydisperse granular materials, but not
  under strictly zero-gravity conditions.}\BibitemShut {Stop}%
\end{thebibliography}

\end{document}